\documentclass[oldversion]{aa}
\usepackage{graphicx}
\usepackage{txfonts}
\usepackage{natbib}
\usepackage{wrapfig}
\def\na1{Na{\sc I}}

\def\msun{M$_\odot$}

\def\fint{ergs\,cm$^{-2}$\,s$^{-1}$}
\def\lum{ergs\,s$^{-1}$}

\def\sds{SDSS\,J120615.73+510047.0}
\def\jxx{SDSS\,J103100.6+202832.2}
\def\kmps{km\,s$^{-1}$}

\begin{document}

   \title{Post common envelope binaries from the SDSS. VI. \sds: a new low
   accretion rate magnetic binary} 
   \author{A.D. Schwope\inst{1}
     \and A. Nebot Gomez-Moran\inst{1}
     \and M.R. Schreiber\inst{2}
     \and B.T. G\"ansicke\inst{3}
}

   \institute{Astrophysikalisches Institut Potsdam,
     An der Sternwarte 16, 14482 Potsdam, Germany\\
     \email{aschwope@aip.de}
     \and
     Departamento de F\'isica y Astronom\'ia, Universidad de Valpara\'iso, Avenida
     Gran Bretana 1111, Valpara\'iso, Chile 
     \and 
     Department of Physics, University of Warwick, Coventry CV4 7AL, UK
      }

   \date{Received ; accepted }
\abstract{We report the discovery of the ninth pre-polar, consisting of a
  late-type ZAMS secondary and a magnetic white dwarf. 
  The white dwarf accretes at extreme low rate, $\dot{M} \sim
  10^{-14}$\,\msun\,yr$^{-1}$, from the wind of the companion donor star. 
  The source was found in our systematic search for WD/MS binaries within
  SDSS/SEGUE. Based on seven Sloan-spectra we estimate a binary period of
  $\sim$\,200, 230, or 270\,min. The UV to IR spectral energy distribution
  was decomposed into a dM3-dM4 ZAMS secondary and a cool white dwarf,
  $\sim$9000\,K, which consistently imply a distance between 360 and 420\,pc. 
  The optical spectrum displays one pronounced cyclotron hump, likely
  originating from a low-temperature plasma, $\sim$1\,keV, in a field of
  108\,MG. We comment on the evolutionary link between polars and pre-polars.
}

\titlerunning{A new magnetic binary from SEGUE}
\maketitle

\section{Introduction}

Ten years ago \cite{reimersetal99} discovered a white-dwarf/main
sequence binary with a very peculiar emission line, while inspecting spectra
of quasar candidates in the HQS objective prism survey.
The line turned out to be the third harmonic of a cyclotron fundamental
emitted by a low-density plasma in a system which was regarded a magnetic
cataclysmic variable (AM Herculis star or a polar) in a persistent low
state. Shortly thereafter, \cite{reimershagen00} found a second system
with very similar properties, which led \cite{schwopeetal02-1} to coin them
LARPs, Low-Accretion Rate Polars. 

In the ensuing years
six further objects of this kind were uncovered in the SDSS 
\citep{schmidtetal05, schmidtetal07} which share
the following properties. They host active late-type main-sequence stars and
accreting, cool, magnetic white dwarfs. All display pronounced
cyclotron spectra originating from low-density plasmas 
on the white dwarfs.
The accretion rates are of order $10^{-13}$\,\msun yr$^{-1}$, orders of
magnitude below the rates expected for polars at the given orbital
periods, which are about $10^{-10}$\,\msun yr$^{-1}$. 
The low accretion rates appeared to be constant over years
\citep{schwarzetal01} and were found to be consistent with the mass loss rate
of the active secondary \citep{schwopeetal02-1}. 

The low accretion rates together with  the system parameters of some
well-studied systems suggested a scenario of underfilling secondaries and no
Roche-lobe accretion at all.
Hence, their class name, referring to them as polars, seems to be a
misnomer.  

The class is still very small, most of their members were serendipituously
found as quasar candidates due to their unusual colors or their broad
(cyclotron) emission lines. Here we report the first detection of such an
object in a project
targeting white-dwarf/main-sequence (WDMS) binaries spectroscopically
following a multi-colour photometric selection process within SDSS/SEGUE
\citep{schreiberetal07} \citep[see also][for a more comprehensive description
of the survey]{rebassaetal07, anebotetal09, mrschreiberetal09, yannyetal09}.

Within SDSS/SEGUE
\citep[for a technical description of the SDSS survey
see][]{fukugitaetal96,gunnetal98, gunnetal06, hoggetal01, ivezicetal04,
  pieretal03, smithetal02, stoughtonetal02, tuckeretal06, yorketal00}
we obtained 533 spectra from a multi-color selected
sample designed to find WDMS binaries with cool white dwarfs and
late-type secondaries. While routinely attempting a spectral
decomposition into its stellar constituents, \sds\ attracted more
interest due to a prominent spectral hump on top of the blue continuum
of a suspected white dwarf. We thus retrieved not only the mean SDSS/SEGUE
spectrum but also the seven individual spectra from the SDSS data base. 
As result of the more comprehensive analysis  we
are confident that \sds\ is the ninth member of the elusive class of
WDMS binaries with a magnetic white dwarf.

We present the system parameters as
derived from the Sloan archive and discuss the class properties.

\begin{figure}
\resizebox{\hsize}{!}{
\includegraphics[bbllx=51pt,bburx=538pt,bblly=49pt,bbury=762pt,angle=-90,clip=]{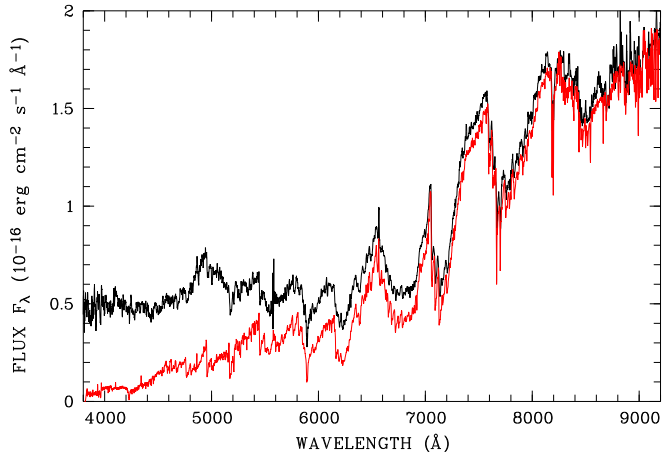}
}
\caption{Mean SDSS spectrum of \sds.
The spectrum was boxcar filtered over three pixels. The spectrum shown in red 
is a dM4-template spectrum constructed from SDSS data.
\label{f:spec}
}
\end{figure}

\section{SDSS observations and analysis} 
With $ugriz$ magnitudes of $20.58,     19.90,   19.32,   18.14,   17.39$ \sds\
was just slightly brighter than our chosen limit for SDSS-spectroscopy, $g=20$
\citep{schreiberetal07}.

\begin{table}[t]
\caption{Time of mid-integration of the seven sub-spectra obtained March
  9/10, 2008, phases according to Eq.~\ref{e:eph} and radial velocity of the
  NaI doublet. The 1$\sigma$ measurement uncertainty is 9\,\kmps.} 
\begin{tabular}{rccr}
\hline
Seq. & Time & Phase & Velocity \\
\#&(HJD) & & (\kmps) \\
\hline
1 & 2454536.246827 & 0.77 & $-300$\\
2 & 2454536.328123 & 0.36 & $ 264$\\
3 & 2454537.220600 & 0.86 & $-188$\\
4 & 2454537.239975 & 0.00 & $  49$\\
5 & 2454537.259523 & 0.14 & $ 195$\\
6 & 2454537.284940 & 0.33 & $ 227$\\
7 & 2454537.299859 & 0.43 & $ 151$\\
\hline
\label{t:obs}
\end{tabular}
\end{table}
 
\begin{figure}
\resizebox{\hsize}{!}{
\includegraphics[bbllx=57pt,bburx=537pt,bblly=56pt,bbury=754pt,angle=-90,clip=]{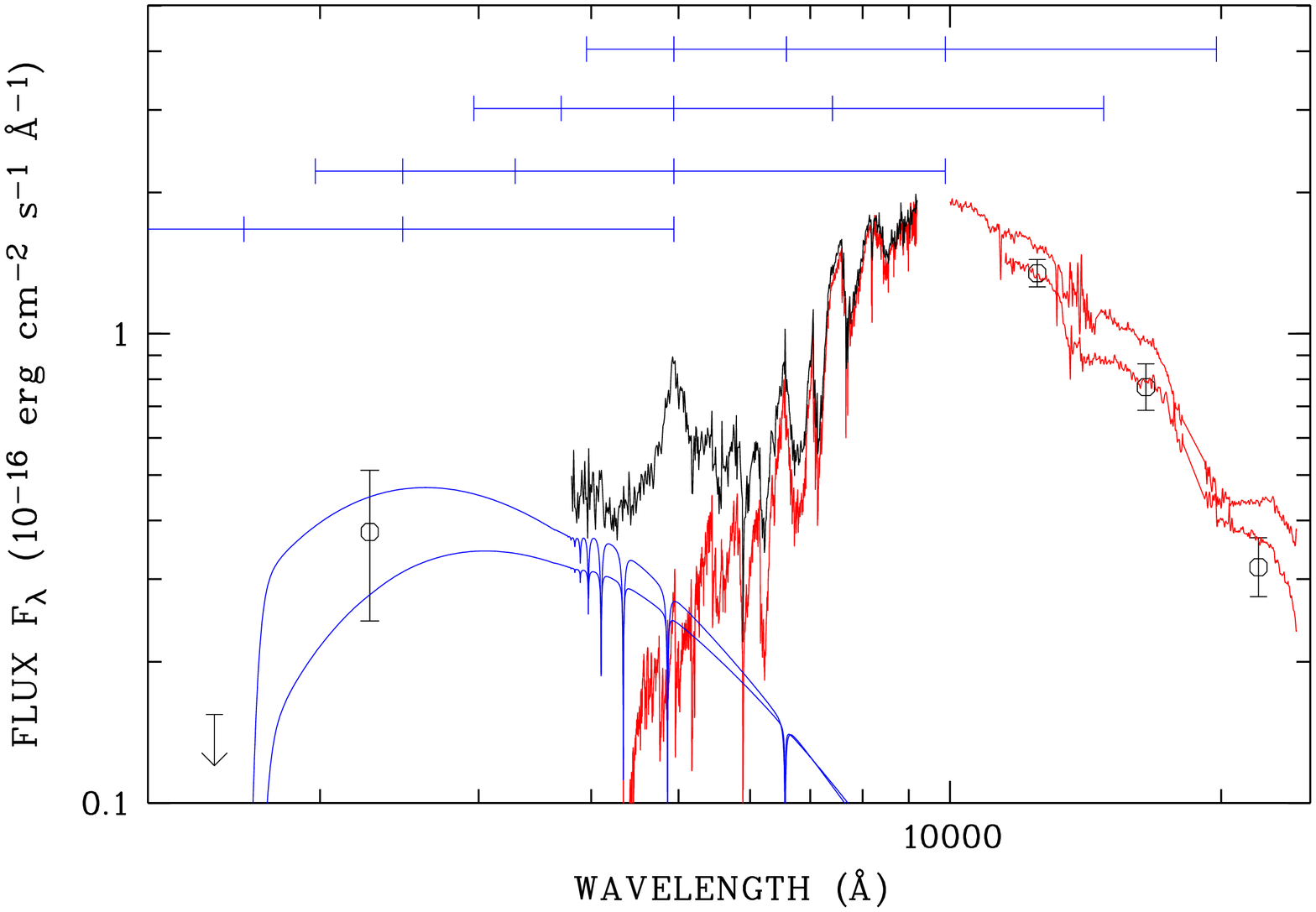}
}
\caption{Spectral energy distribution of \sds. In black are shown data
  obtained from 2MASS (JHK photometry), SDSS (optical spectrum), and GALEX (UV
  photometry). Spectral templates for the secondary obtained from the SDSS
  (optical) and from \citet{leggettetal00} are shown in red. DA model spectra
  for 8000\,K and 9000\,K are shown in blue. The expected locations of the
  first five harmonics in fields of 216, 108, 72, and 54\,MG are indicated by
  short ticks in the upper part of the diagram.
\label{f:sed}
}
\end{figure}

\subsection{Average SDSS spectrum and the spectral energy distribution}

The average spectrum of \sds\ published in SDSS-DR7
\citep{abazijianetal09} is shown in Fig.~\ref{f:spec}.
It is dominated by the late-type secondary star whose spectral features
are best reflected with a dM4 ZAMS template spectrum.
The distance to \sds\ derived from a scaled dM4
template spectrum is $d_{\rm MS} = 420\pm120$\,pc.
\sds\ was detected also with 2MASS and the IR colours and the IR-to-optical
spectral energy distribution suggest a slightly earlier spectral type M3.
Figure~\ref{f:sed} shows the SDSS spectrum and also contains a suitably scaled
M4 spectral template in the optical, 
and IR spectra of LHS57 (M4) and LHS54 (M3), both taken
from Legget's spectral archive \citep{leggettetal00} scaled to the same 
optical brightness. 
While an M3-type secondary and corresponding larger distance is
compliant with the data, we nevertheless use an M4 template for the spectral 
deconvolution of the SDSS-spectra.

The field of \sds\ was observed for 105\,s with GALEX, and
\sds\ was detected in the NUV channel but not detected in the FUV channel. 
The GALEX fluxes (1$\sigma$ upper limit for the FUV channel) are included in
Fig.~\ref{f:sed}, too. 

The residual SDSS-spectrum after subtraction of the M-star template can be
described by a blue featureless continuum with a broad blue hump
superimposed. The absence of any absorption line suggests a
classification as DC white dwarf (but see the discussion below).
Nevertheless, we use as first order approximation for the white-dwarf
spectral flux a DA model atmosphere. The optical to ultra-violet
spectral energy distribution is well reflected with a low-temperature
white dwarf of 8000 -- 9000\,K (see Fig.~\ref{f:sed}). Assuming
9000\,K and an average white dwarf of $M_\mathrm{wd}=0.6$\,\msun\ and
$R_\mathrm{wd}=8.3\times10^{8}$\,cm the distance to
\sds\ is $d_{\rm WD}=360$\,pc, in good agreement with $d_{\rm MS}$.

\begin{figure}
\resizebox{\hsize}{!}{
\includegraphics[bbllx=56pt,bburx=527pt,bblly=73pt,bbury=750pt,angle=-90,clip=]{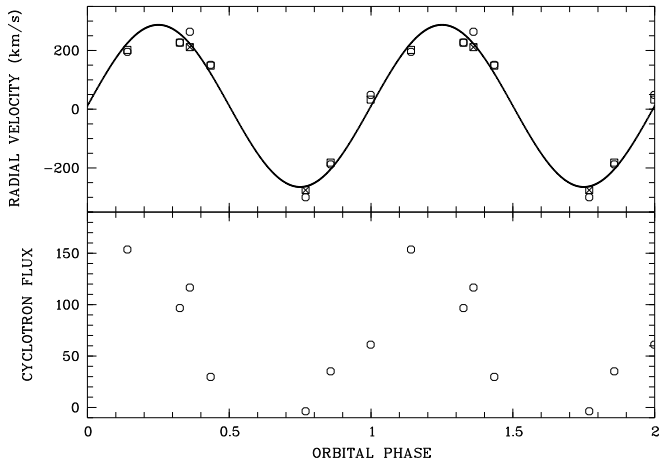}
}
\caption{{\it (top panel)} Radial velocities of H$\alpha$ emission (squares)
  and Na absorption lines (circles). The two data points from the first night
  are crossed. The continuous line is a sine fit to the Na radial velocities. 
{\it (bottom panel)} Integrated cyclotron flux of the 4940\,$\AA$\,spectral
  hump as a function of orbital phase. Flux units are
  $10^{-15}$\,erg\,cm$^{-2}$\,s$^{-1}$. 
  The centroid of the bright orbital hump
  at phase 0.15 indicates the likely azimuth of the accretion spot.
\label{f:rv}
}
\end{figure}

\subsection{Line variability}
The SDSS/SEGUE spectrum of \sds\ published via SDSS-DR7 is the 
average of seven sub-spectra with exposure 15 min each obtained during
  the two nights of March 9 and 10, 2008, respectively (see Tab~\ref{t:obs}
  for exact dates). 
We retrieved the individual
spectra from the Sloan database to search for spectral and photometric
variability. 
 
H$\alpha$ emission and NaI absorption lines were used to search for radial
velocity variations by fitting Gaussians to the observed data. The double
Gaussian for the NaI lines was fitted with fixed separation between the two
lines and with the same width of both lines. The line fluxes of both,
H$\alpha$ emission and NaI absorption lines, show insignificant
variability. Both line features are resolved with measured FWHM of
$\sim$5\,\AA\ (H$\alpha$) and 6.5\,\AA\ (NaI). Again, variability of 
the width is insignificant in the present data.

Both line features display pronounced radial velocity variations with
peak-to-peak amplitude of 500\,\kmps\ (H$\alpha$) and 560\,\kmps (NaI),
respectively. The sequence of five spectra obtained March 10 rule out 
any period below 3 hours. A Lomb-Scargle periodogram of the radial velocities
gives almost equal power at 5.28, 6.30, and 7.32 cycles per day (corresponding
to periods of 270\,min, 230\,min, and 200\,min, respectively). Without loss of
generality we assume the shortest period to derive a spectroscopic ephemeris
and to assign binary phases to individual spectra in this paper. 
The ephemeris of the blue-to-red zero crossing of the NaI lines thus derived 
is 
\begin{equation}
\label{e:eph}
\mbox{HJD} = 2454537.2569(1) + E \times 0.1366(1)
\end{equation}
The quoted 1$\sigma$ uncertainties were derived from a sine fit to the NaI
radial velocities. The radial velocity curves of H$\alpha$ and NaI lines are
shown in Fig.~\ref{f:rv}. Both species display a common sine-like radial
velocity curve, suggesting that H$\alpha$ is due to stellar activity and not
due to accretion. 

\begin{figure}
\resizebox{\hsize}{!}{
\includegraphics[bbllx=38pt,bburx=490pt,bblly=100pt,bbury=767pt,clip=]{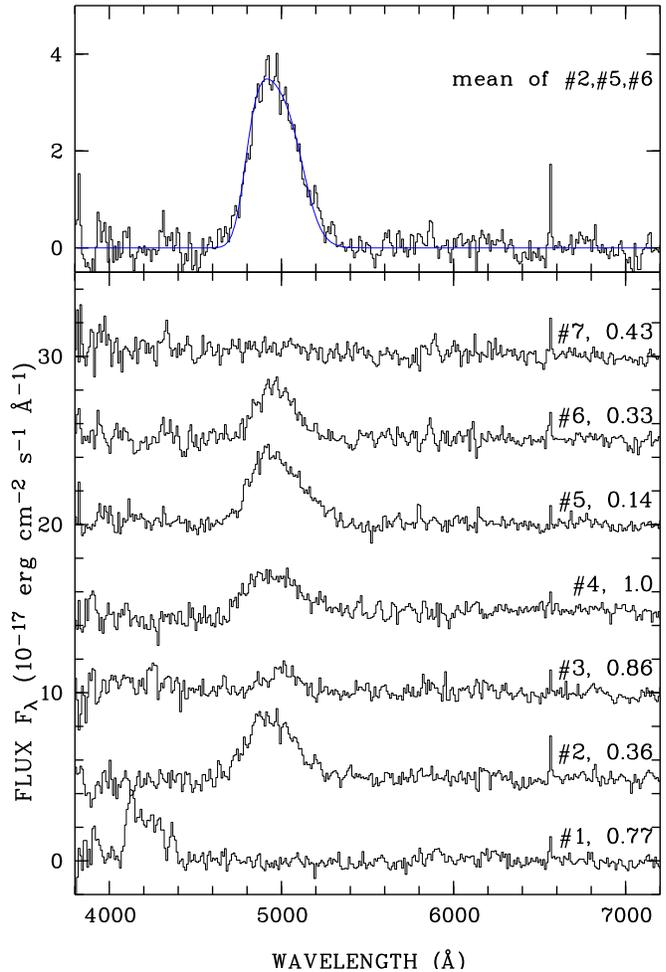}
}
\caption{{\it (lower panel)} 
Cyclotron spectra of \sds\ generated by subtracting a ZAMS M4
  template spectrum and a smooth blue continuum. The individual spectra were
  plotted with an offset of 5 flux units. Phases are indicated according to the
  ephemeris given in Eq.~\ref{e:eph};
{\it (upper panel)} Mean of spectra with sequence numbers as indicated, compared
  with a cyclotron model emitted in a 1 keV plasma at 108\,MG.
\label{f:cyc}
}
\end{figure}

\subsection{Continuum variability -- the cyclotron spectrum}
The contribution of the assumed M4 ZAMS secondary star was subtracted
from the seven 
individual spectra. These were then summed and a smoothed continuum was
defined by a polynomial fit to 10 points in wavelength regions omitting the
spectral hump at 4940\,\AA. Again, the
same smooth continuum was subtracted from all individual spectra. The
residuals are shown in original time sequence running from bottom to top in
Fig.~\ref{f:cyc}. Phases according to Eq.~\ref{e:eph} are indicated (see also
Tab.~\ref{t:obs} for epochs and phases of individual spectra). The only
remaining significant feature in the spectra apart from H$\alpha$ is the broad
hump which we interpret as a cyclotron harmonic in emission. The blue feature
centered on 4200\,\AA\ in spectrum \#1 cannot be classified. 
It may be a cyclotron line from a second region, but the flux changes rather
abruptly and not as smooth as seen in other cyclotron lines. 
It may therefore well be an artefact of the observation/reduction. 
Identification of this feature being either physical or instrumental 
needs further phase-resolved spectroscopy. 

Although we could identify just one hump in the spectra we think there is
little doubt about the nature of the feature being of cyclotron origin. The
interpretation rests on the similarity to other systems, its variability
pattern (again similar to other objects of this class), and the success of our
modeling (see below). Final confirmation
needs the identification of at least one neighbouring cyclotron harmonic or the
detection of a polarized signal from the observed hump at 4940\,\AA.

Assuming the cyclotron interpretation being correct, some basic parameters 
of the radiation source can be derived/constrained, in the first place the
strength of the magnetic field, $B$. A proper determination 
would require to measure the separation between two adjacent harmonics, 
but the isolated harmonic already provides some strong constraints.

Short vertical ticks in 
Fig.~\ref{f:sed} indicate the expected positions of cyclotron harmonics 
$1-5$, would
the observed one be the 1$^{\rm st}$, 2$^{\rm nd}$, 3$^{\rm rd}$, or the
4$^{\rm th}$ harmonic of the cyclotron fundamental in a field of $B = 54.2, 
72.3, 108.4,$ or $216.8$\,MG, respectively. 
The fact that the one observed hump is isolated and well separated
from its next (non-detected) neighbour 
excludes any field as low as 54 MG (observed hump would be 
harmonic number $n=4$). 
A field strength of 72\,MG ($n=3$) seems unlikely too (but not completely
excluded), since the 2$^{\rm nd}$
harmonic at $7500 \AA$ should be detectable. We remain with mainly two 
alternatives, $B = 108$ or 216 MG, and we are inclined to accept 
the former for the simple reason that the latter appears to be 
extraordinarily high.

The plasma temperature in the accretion region is low because the line does
not show any significant wavelength shift although the large photometric
variability suggests that the viewing angle to the field is strongly
variable. Furthermore, the line is rather narrow with $350 \AA$ (FWHM).
Cyclotron spectra are determined by the magnetic field strength, the 
temperature, the viewing angle with respect to the field and the optical 
depth of the plasma at the given frequency \citep{schwope90}.
Assuming optically thin radiation and a moderate viewing angle of 60\degr, the
plasma temperature can be crudely estimated. In Fig.~\ref{f:cyc} a model for
the second harmonic in a 1 keV plasma at $B=108$\,MG is shown, which nicely
represents the data (mean of the three brightest spectra). 

The peak
integrated flux of the cyclotron hump among the seven spectra is 
$F_{\rm cyc} = 1.5 \times 10^{-15}$\,\fint (see Fig.~\ref{f:rv}). 
If we assume a bolometric correction factor 2, 
the cyclotron luminosity is $L \simeq 2 F_{\rm cyc} 2\pi d^2 = 
3\times 10^{28} (d/400\mbox{pc})^2$\,\lum. The implied mass accretion rate 
is estimated by equating the cyclotron luminosity to the accretion luminosity,
$\dot{M} \simeq 10^{-14}$\,\msun yr$^{-1}$. Although these numbers are 
uncertain by factors, it is clear, that \sds\ belongs to the class 
of low-luminosity, detached magnetic white-dwarf/main-sequence binaries 
found earlier in the HQS and the SDSS. The accretion rate is
consistent with  wind-accretion from  the active secondary.

In Fig.~\ref{f:rv} (lower panel) the optical cyclotron light curve is
displayed as a function of the orbital phase. If one assumes 
spin-orbit locking of the white dwarf \sds\ has a cyclotron bright phase
lasting for about 0.6 orbital cycles, centered on phase $\sim$0.1. 
This phasing suggests that the accretion region 
is trailing the secondary by about 35 degrees in phase. This orientation 
is different from that which one typically finds in the polars, where the
accreting pole is on the leading side \citep{cropper90}. 
Spin-orbit synchronism is observed for all well-observed objects of this class
but needs to be proven for the new object by further photometry and
spectroscopy.  

\begin{figure}
\resizebox{\hsize}{!}{
\includegraphics[bbllx=85pt,bburx=505pt,bblly=92pt,bbury=650pt,clip=]{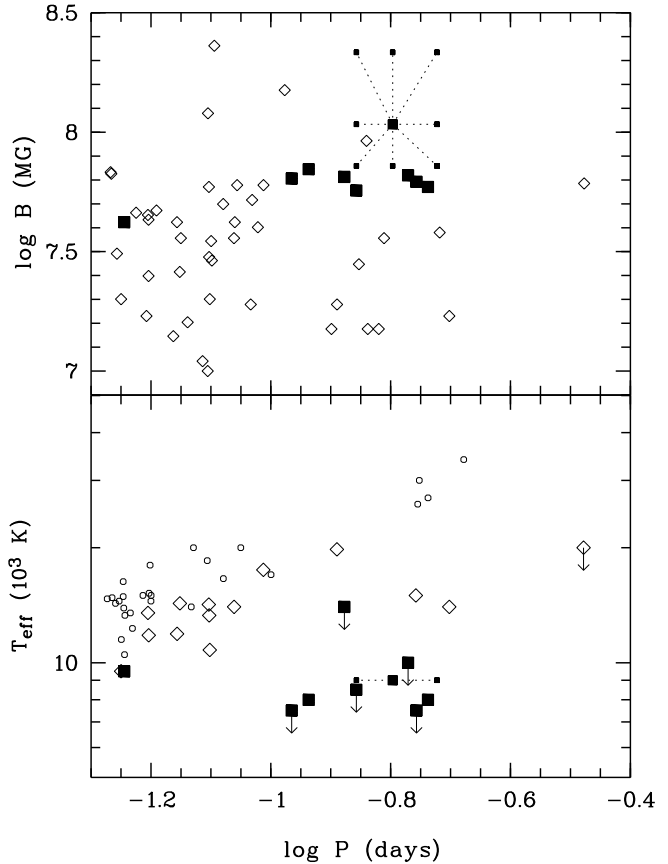}
}
\caption{{\it(top)} Magnetic field versus orbital period of accreting close
  binaries containg a magnetic white dwarf and a main-sequence secondary. Open
  symbols denote binaries accreting via Roche-lobe overflow, the Polars. Filled
  symbols denote PREPs, filled symbols connected by dotted lines indicate
  possible locations of \sds.
{\it(bottom)} Effective temperature of accreting magnetic and non-magnetic
  white dwarfs. Open circles indicate nonmagnetic white dwarfs, otherwise
  symbols are as above. 
\label{f:dist}
}
\end{figure}

\section{Discussion and conclusions}
We have identified \sds\ as a new member of the class of close, but still 
detached, binaries containing a late-type ZAMS dwarf and a cool 
magnetic white dwarf accreting at extreme low rate from a stellar wind of the
secondary star. 

{\it How to name these objects?}  When less than a handful of those
were known they were recognized as Low Accretion Rate Polars, or
shortly LARPs \citep{schwopeetal02-1}, an acronym meanwhile widely
used. The implication of this naming convention was that those objects
are just ordinary polars which entered an extended low state. This is
true for e.g.~EQ Cet (aka RBS206), which was considered a LARP because
of its peculiar cyclotron spectrum \citep{schwopeetal99}. However, it
became later clear \citep{schmidtetal05,schwarzetal01,vogeletal07}
that the secondaries of most LARPs seem to be Roche-lobe underfilling
and hence accretion via Roche-lobe overflow, the defining criterion of
a cataclysmic variable, can not occur and has not occurred in the past. 
Hence, they are pre-cataclysmic
binaries or more precisely, pre-polars, henceforth PREPs. Objects such
as EQ\,Cet showing intermittent high accretion states
\citep{schwopeetal02-2} are consequently not members of this class.

We now discuss the system parameters of the ninth member and some of
the class parameters of the PREPs.  Based on the Sloan data alone, we
could estimate the orbital period, the magnetic field strength of the
white dwarf, the spectral type of the secondary, and the distance to
the system. GALEX and 2MASS data were helpful to refine the stellar
parameters.

The magnetic field of the white dwarf was identified via a remarkable
spectral hump at 4940\,\AA\ identified as the likely 2nd cyclotron
harmonic in a low-temperature plasma with $B = 108$\,MG.
Interestingly, none of the now nine members of this class displayed
photospheric magnetism via Zeeman-split absorption lines, a likely
combined effect of their faintness, their coolness and the expected
large Zeeman spread in fields as high as 60 -- 100\,MG. If confirmed,
the field of 108 MG (or even 216\,MG)
in \sds\ extends the parameter range of PREPs to
higher magnetic fields.

The white dwarf in \sds\ is cool, $T_{\rm eff} < 10000$\,K, 
a common property of the class.

Polars and PREPs are thought to contain the same types of stars, magnetic
white dwarfs and late-type ZAMS secondaries. While
polars are accreting via Roche-lobe overflow with an estimated duty cycle of
about $\sim$50\% \citep{hessmanetal00}, the latter are thought to
accrete from the stellar wind of the secondary via a magnetic siphon
\citep{schwopeetal02-1,webbinkwick05,schmidtetal05,vogeletal07}. 

A comparison of the magnetic fields, the white-dwarf effective temperatures
and the orbital periods between the polars and the PREPs is instructive
and shown in Fig.~\ref{f:dist}. 
Values of the magnetic fields are based on our own
compilation whereas the white-dwarf effective temperatures of polars and
non-magnetic CVs are from the recent compilation by
\citet{townsleygaensicke09}. 
Aan upper limit temperature of 20000\,K was added for
V1309\,Ori \citep{staudeetal01}.
Effective temperatures 
for PREPs are from \citet{schmidtetal05,schmidtetal07,vogeletal07} and this
work. 

\citet{araujo-betancoretal05-2} and \cite{townsleygaensicke09} have discussed
the distribution of $T_{\rm eff}$ as 
a function of orbital period for non-magnetic and magnetic CVs (polars). They 
conclude that the temperature contrast between
magnetic and non-magnetic white dwarfs which exists at any given period must
arise from a difference in the time-averaged mass accretion rate,
$\langle\dot{M}\rangle$. This is consistent with the suggestion that
polars have lower angular momentum loss rates due to a reduced
efficieny of magnetic braking \citep{wickramasinghe+wu94-1}. 

Fig.~\ref{f:dist} shows that the two classes containing a magnetic white
dwarf, polars and PREPs, also separate rather well in temperature and magnetic
field. 
All but one, SDSS\,J103100.6+202832.2 \citep{schmidtetal07}, 
PREPs have high 
magnetic fields and long orbital periods, most of them are found above 
the cataclysmic variable period gap. For MCV standards the objects
thus appear mildly young. On the other hand, all have low white-dwarf
effective temperatures which one might be tempted to regard as age indicator.

\citet{townsleygaensicke09} have shown that
compressional heating in polars is less efficient compared to non-magnetic CVs
but still important. The separation between polars and PREPs in the $(T_{\rm
  eff}, P)$ plane thus confirms that they are distinct classes, the WDs in
polars being systematically hotter due to compressional heating. There is 
one interesting exception, \jxx, at same temperature and
orbital period as EF Eri, a well-studied polar with a cool white dwarf
\citep{schwopeetal07}. The striking resemblance between EF Eri and \jxx\
  was 
discussed already by \citet{schmidtetal07} in their discovery paper. 
As a consequence of accretion heating,
$T_{\rm eff}$ will not serve as age indicator for polars, but likely for
PREPs due to their very low accretion rates.

Figure~\ref{f:dist} shows that magnetic fields in PREPs are clustering around
$60-65$\,MG, with one clear exception, \jxx\ at 42\,MG, and another likely
exception, \sds\ at 108\,MG. While
it is tempting to speculate about an evolutionary link between high magnetic
fields on the one hand and long orbital periods/cool white dwarfs on the other
hand, care must be taken about selection effects. The tenous, low-temperature
plasmas in PREPs have radiative power only in the first few cyclotron
harmonics. The cyclotron fundamental is likely optically thick (although not
observed yet), and the fourth harmonic in several of the known cases is optically
thin and rather difficult to detect. A straightforward detection by spectroscopic
means as done in the past is more or less easily feasable, if the second or
third harmonic lies between $4500\,\AA$ and $\sim$8000\,$\AA$, corresponding
to $B = 45 \dots 120$\,MG. The object \jxx\ is slightly below this range. Its 
fortunate discovery was possible due to its rather large power in harmonics
higher than the third while \sds\ is possibly close to the upper limit. 

Low-field PREPs, i.e.~those with cyclotron spectra in the infrared, could be
recognized as such, if Zeeman-split photospheric lines could be identified. at
least hypothetically. But even the known PREPs don't show any easily
identifiable Zeeman feature in their flux spectra\footnote{which is due to their low
temperatures, hence their faintness and the small equivalent widths of the
lines} contrary to the white dwarfs in low-state polars which are hotter on
average, which makes the
identification of low-field PREPs practically very difficult. Low-field PREPs
could be hidden among the WDMS-systems with apparent DC white dwarfs and would
require either spectro-polarimetry to be identified or flux spectra with
higher signal to noise.

Whether the distribution of field strength between polars and PREPs is truly
different or not, remains unanswered for the time being, but their period
distributions are very different and less affected by observational selection
effects.

One explanation for the discrepant period distributions originally formulated
by \citet{webbinkwick05} and later by \cite{schmidtetal05}
assumes early synchronization of the white dwarfs in PREPs
due to their (high?) magnetic field. Early synchronization was supposed to
slow down binary evolution by an effective reduction or even 
complete cessation of magnetic braking above the orbital period gap.
As a result, comparatively cool white dwarfs are observed even at long orbital
periods. This scenario involves different angular momentum loss (AML) rates for
polars and PREPs.

An alternative is to assume the same AML for both classes, the
separation in $P$ then implies that PREPs are young compared to polars. 
Observationally both scenarios are difficult to discern, since $T_{\rm eff}$ cannot 
be used as age indicator.

{\it Are the PREPs the missing detached magnetic white dwarf/main sequence
  binaries?} 

\citet{liebertetal05} noted the absence of non-accreting magnetic  white
dwarfs (MWDs) among the more than 1200 WDMS binaries compiled by
\citet{silvestrietal07} and various other samples 
\citep{marsh00-1, ritter+kolb03-1,
schreiber+gaensicke03-1,morales-ruedaetal05-1, shimanskyetal06-1}.  
MWDs are clearly
underrepresented in current WDMS samples compared to the $\sim$10\% fraction in 
the field \citep{liebertetal03}. As a first possible solution to this fact,
\citet{liebertetal05} suggested that, analogous to single magnetic
white dwarfs, the magnetic white dwarfs in WDMS binaries are more
massive, and therefore smaller and less luminous compared to those in 
non-magnetic WDMS binaries which could result in an observational 
bias against their detection. However, \citet{silvestrietal07} 
convincingly demonstrate that such systems should be easily identified, if
they were present in the SDSS data base. 

Stimulated by the recent observational studies, \citet{toutetal08}
propose a rather radical shift in paradigm for the formation of highly
magnetic white dwarfs, suggesting that the formation of high-field
white dwarfs in general is tightly related to the evolution through a
common envelope phase. In that scenario, single high-field magnetic
white dwarfs are the results of mergers during the common envelope,
explaining the higher average mass of magnetic white dwarfs compared
to non-magnetic white dwarfs. Those systems that avoid merging leave
the common envelope as pre-polars with such short orbital periods that
their white dwarfs can capture some of the wind of the companion star
(in other words, they leave the CE as PREPs).

However, the cold temperatures of the white dwarfs in PREPs provide us with 
strong lower limits on their post common envelope lifetimes 
of the order of several hundred Myrs. 
Unfortunately, as the stellar masses of the PREPs 
remain unknown, we can neither determine the detailed cooling ages nor can we 
reconstruct their detailed post-CE evolution and, hence, cannot answer the 
question whether they could have been wind-accreting systems already 
shortly after the common envelope phase. However, 
the fact that all PREPs contain cold white dwarfs, i.e.~are old post common 
envelope binaries, implies that the appealing idea 
presented by \citet{toutetal08} does 
not solve the missing MWD problem but requires to rephrase it:  
{\it If PREPs are indeed the progenitors of polars, where are the 
progenitors of PREPS, i.e.~WDMS systems containing hot magnetic white dwarfs?}

\begin{acknowledgements}
We thank our referee, G.D. Schmidt, for valuable comments and suggestions
which helped to improve the manuscript.
We also thank Justus Vogel for comments on an earlier version of the
manuscript. ANGM was supported by the Deutsches Zentrum f\"ur Luft- und
Raumfahrt (DLR) GmbH under contract No.~50OR0404.  

Funding for the SDSS and SDSS-II has been provided by the Alfred P. Sloan
Foundation, the Participating Institutions, the National Science Foundation,
the U.S. Department of Energy, the National Aeronautics and Space
Administration, the Japanese Monbukagakusho, the Max Planck Society, and the
Higher Education Funding Council for England. The SDSS Web Site is
http://www.sdss.org/. 

The SDSS is managed by the Astrophysical Research Consortium for the
Participating Institutions. The Participating Institutions are the American
Museum of Natural History, Astrophysical Institute Potsdam, University of
Basel, University of Cambridge, Case Western Reserve University, University of
Chicago, Drexel University, Fermilab, the Institute for Advanced Study, the
Japan Participation Group, Johns Hopkins University, the Joint Institute for
Nuclear Astrophysics, the Kavli Institute for Particle Astrophysics and
Cosmology, the Korean Scientist Group, the Chinese Academy of Sciences
(LAMOST), Los Alamos National Laboratory, the Max-Planck-Institute for
Astronomy (MPIA), the Max-Planck-Institute for Astrophysics (MPA), New Mexico
State University, Ohio State University, University of Pittsburgh, University
of Portsmouth, Princeton University, the United States Naval Observatory, and
the University of Washington.  

\end{acknowledgements}

\bibliographystyle{aa}
\bibliography{references}
\end{document}